\documentclass[preprint,pra,aps,nofootinbib,superscriptaddress]{revtex4-2}
\usepackage[mathscr]{euscript}
\usepackage{graphicx}
\usepackage{float,epsfig}
\usepackage{dcolumn}
\usepackage{bm}
\usepackage{graphicx}
\usepackage{amsmath,amssymb,amsthm}
\usepackage{subfigure}
\usepackage{color}
\usepackage[colorlinks=true,linkcolor=blue]{hyperref}
\usepackage[normalem]{ulem}
\usepackage[makeroom]{cancel}
\usepackage{caption}
\usepackage{subcaption}
\usepackage{textcomp}
\usepackage{ragged2e}
\usepackage{mathalfa}

\newcommand{\bea}{\begin{eqnarray}}
\newcommand{\eea}{\end{eqnarray}}
\newcommand{\beq}{\begin{equation}}
\newcommand{\eeq}{\end{equation}}

\def\/{\over}
\definecolor{purple}{rgb}{0.8,0,1}

\begin{document}

\title{Optical branched flow in nonlocal nonlinear medium}

\author{Tongxun Zhao}
\affiliation{School of Physics and Astronomy, Shanghai Jiao Tong University, Shanghai 200240, China}

\author{Yudian Wang}
\affiliation{School of Physics and Astronomy, Shanghai Jiao Tong University, Shanghai 200240, China}

\author{Ruihan Peng}
\affiliation{School of Physics and Astronomy, Shanghai Jiao Tong University, Shanghai 200240, China}

\author{Peng Wang}
\affiliation{School of Physics and Astronomy, Shanghai Jiao Tong University, Shanghai 200240, China}

\author{Fangwei Ye} 
\email{fangweiye@sjtu.edu.cn}
\affiliation{School of Physics and Astronomy, Shanghai Jiao Tong University, Shanghai 200240, China}
\affiliation{School of Physics, Chengdu University of Technology, Chengdu, China}
\date{\today}


\begin{abstract}
\textbf{When light propagates through a randomly correlated, slowly varying medium, it generates optical branched flow. Previous studies have demonstrated that the self-focusing effect in optical media can accelerate the appearance of the first branching points and sharpen the filaments of branched flow. In this study, we investigate the influence of the nonlocality of the nonlinear response on branched flow. We find that, due to its averaging effect, as the range of nonlocality increases, the first branching point shifts to a greater distance, and the flow structures broaden, thus nonlocality  ultimately restores the branched flow to its linear condition. We have developed a semi-analytical formula and confirmed the screening of the self-focusing effect on branching flow by nonlocality.
}

\end{abstract}

\maketitle

\section*{1 Introduction} 
When a  wave passes through a disordered, slowly varying potential, it undergoes multiple small-angle refractions, splitting into several thin filamentary beams. These filaments further divide, forming numerous branches, resulting in significant intensity fluctuations across the propagation cross-section, which is known as branched flow. The phenomenon of branched flow was first discovered for matter waves in experiments of two-dimensional electron gases\cite{topinka2001coherent,bram2018stable,aidala2007imaging,jura2007unexpected,fratus2021signatures,maryenko2012branching,kozikov2013interference,liu2013stability}, where it was observed that electron flow split into several branches of varying thickness, forming a structure resembling the continuous branching of tree limbs. Subsequently, branched flow has been observed in various waves of different natures, including sound waves\cite{wolfson2001stability}, water waves\cite{degueldre2016random,ying2011linear}, and electromagnetic waves\cite{hohmann2010freak,barkhofen2013experimental}.
In 2020, the branched flow for light waves was firstly discovered when a laser beam passed through a soap membrane with non-uniform thickness\cite{patsyk2020observation}, and very recently, it was also observed in nematic liquid crystals with randomly distributed molecular orientation \cite{chang2024electrical,yudynamic}.

Although branched flow was initially introduced for linear systems, where the evolving wave within the material does not alter the property of the potential, the physical realizations of branched flows mentioned above offer opportunities to explore the branched flow in nonlinear contexts, for example, in the presence of optical Kerr nonlinearity in optical settings or electron-electron interactions in electron gases. In these nonlinear regimes, waves can substantially modify the original random potential landscapes, and in turn affecting the wave propagation itself. Thus, it has been revealed that nonlinearity exerts a strong influence on branched flow, reducing the onset distance for the branching occurs through self-focusing nonlinearity and sharpening the flow structures~\cite{green2019branched}, potentially leading to the formation of extreme waves~\cite{mattheakis2016extreme,akhmediev2009extreme,mattheakis2015extreme,zannotti2020caustic} and rouge waves~\cite{dudley2014instabilities,safari2017generation}.

The aforementioned papers on branched flow in nonlinear media focus on the simplest model of a pure Kerr nonlinear medium and do not take into account a potential nonlocal nonlinear response.  The nonlinear response can be spatially nonlocal, meaning that the material’s response at one position is determined not only by the excitation at that particular point but can also be affected by excitation in its neighboring areas. Nonlocality of the nonlinear response is a generic property of various nonlinear material, arising when nonlinearity mechanisms such as carrier diffusion, molecular reorientation, heat transfer, etc., are involved \cite{krolikowski2004modulational}. Typically, nonlocal nonlinear media is characterized by a response function whose characteristic transverse scale determines the degree of nonlocality, denoted as $d$ in the following. In such media, a laser beam with intensity $I$ induces a refractive index change $n$, which is described by the diffusion-type equation $n-d\Delta n=I$, as seen, for example, in liquid crystals~\cite{ye2007enhanced}.

In this study, we investigate the influence of the nonlocal nature of the nonlinear response on branched flow. We find  that while local, self-focusing nonlinearity causes branching to occur earlier and results in sharper flow structures compared to a linear medium, the nonlocal character of the nonlinearity has the opposite effect: As the nonlocality length $d$
increases, the occurrence of branching is delayed until it returns to the position observed in a linear medium. Furthermore, with increasing nonlocality, the structural flows gradually broaden and smoothen. This reversion to the linear scenario due to the nonlocality is attributed to the fact that the nonlocal response is based on the average light intensity within a specific region, thereby reducing the sharpness of the refractive index landscape in the otherwise local, self-focusing nonlinear material. We developed  a ray-tracing model that incorporates the nonlocal nonlinear response, which clearly demonstrates that nonlocality indeed counteracts the effects of local nonlinearity, thereby corroborating our observations in branching dynamics simulations.

\vspace{-0.5em} 
\begin{figure}[H]
     \centering
     \includegraphics[width=1\linewidth]{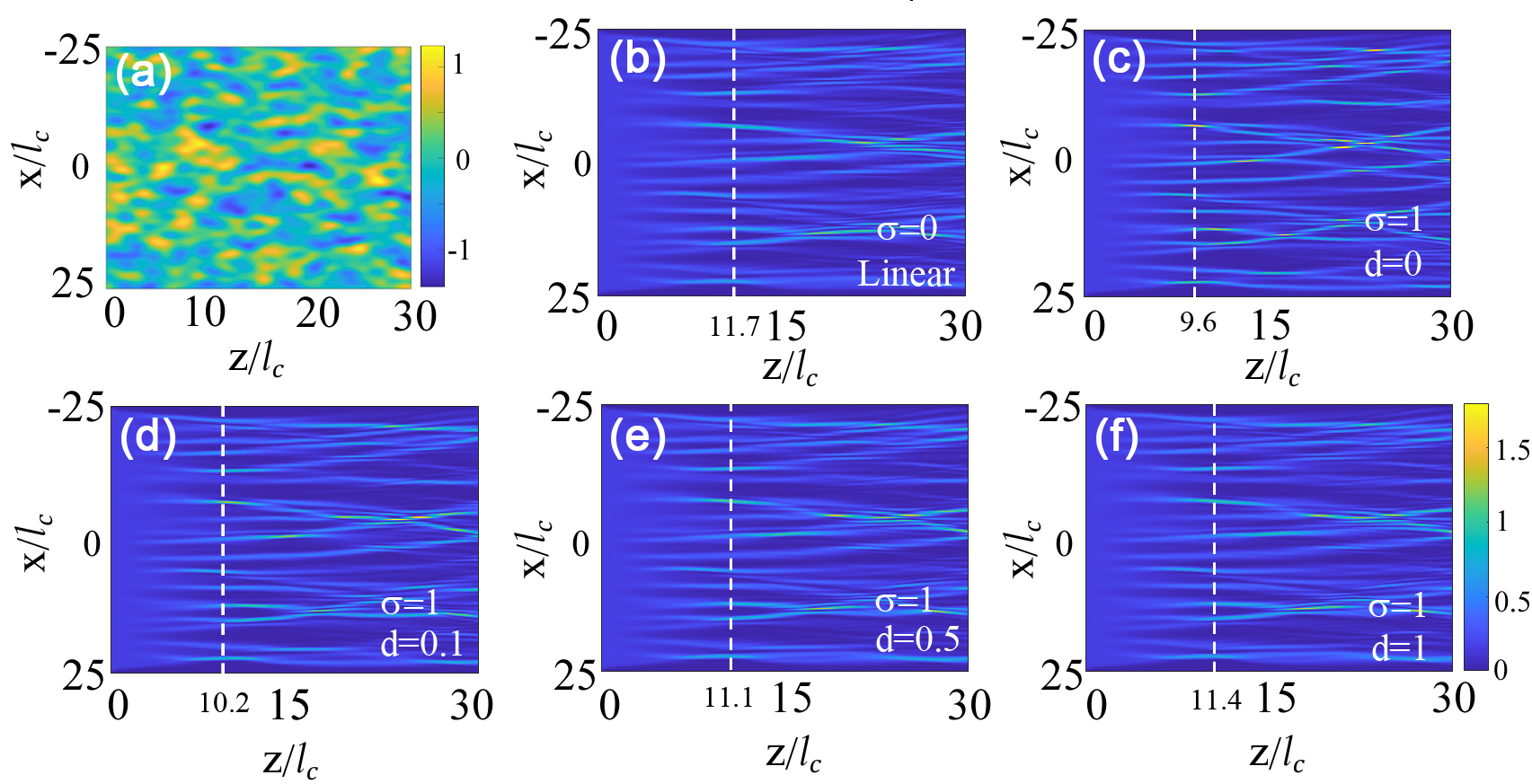}
     \captionsetup{font={stretch=1}}
     \captionsetup{font=footnotesize} 
     \captionsetup{labelfont=bf,name=Fig.}
     \caption{
     {\bf   Propagation of a plane wave through a random potential under various nonlocal nonlinear conditions, with nonlinear coefficient $\sigma$ and degree of nonlocality $d$.
     } 
     (a) shows the landscape of the one specific realization of random potential $V(x,z)$ with $l_c=1$ and $\epsilon=1$. (b) shows the linear propagation result, $\sigma=0$ and (c-f) presents the nonlinear case with $\sigma=1$ and $d$ values of 0,  0.1, 0.5 and 1, respectively. The white dashed lines indicate the z-position of the first branching points, which are calculated using Eq.~(4) by taking the average along the transverse $x$ direction, without averaging over different realizations of random potentials.
    }
     \label{fig1}
 \end{figure}

\section*{2 Model}

Our analysis starts from the propagation of a light beam along the z-axis in a medium with a nonlocal focusing Kerr-type nonlinearity, that is described by the following set of equations for dimensionless complex light field amplitude $\Psi$, and nonlinear change of the refractive index $n$,

\begin{align}
\label{nlse}
   i \frac{\partial \Psi(x, z)}{\partial z}=-\frac{1}{2} \frac{\partial^{2} \Psi(x, z)}{\partial x^{2}}-n \Psi(x, z)-V(x, z) \Psi(x, z)  
\end{align}

\begin{align}
\label{n}
   n(x,z)-d \frac{\partial^{2} n(x,z)}{\partial x^{2}}=\sigma|\Psi(x, z)|^{2} 
\end{align}

Here, $\Psi(x, z)$ represents the complex amplitude of the light wave, and the square of its absolute value, $I=|\Psi(x, z)|^{2}$, corresponds to the intensity of the wave. $x$ and $z$ are the transverse and longitudinal coordinates scaled to the beam width and the diffraction length, respectively. The parameter $d$ represents  the degree of nonlocality of the nonlinear response. It should be noted that, when $d$ approaches $0$, $n$ becomes $n=\sigma|\Psi(x, z)|^{2}$, thereby recovering the local Kerr limit. On the other hand, when $d$ approaches infinity, the system transitions into a strongly nonlocal regime. It is worth mentioning that this diffusion-type nonlocal nonlinear response accuractely describes the nonlinear response of nematic liquid crystals in steady state~\cite{ye2007enhanced}, where the nonlocality degree $d$ is controlled by the applied biasing field $E_\text{b}$.

The function $V(x,z)$ in Eq.~(1) stands for the linear potential, which is assumed to be a smoothly varying random function with respect to $x$ and $z$. This randomness is a prerequisite for the occurrence of the branched flow. To characterize the random potential $V$, we assume  it follows a Gaussian auto-correlation function with a width $l_c$, representing the spatial correlation length of the random potential, and 
an amplitude $\epsilon$, representing the strength of the random potential,

\begin{align}
\label{fp}
  \left \langle f\left ( \Delta x,\Delta z \right )  \right \rangle 
= \left \langle V(x,z)V(x+\bigtriangleup x, z+\bigtriangleup z) \right \rangle  =\varepsilon ^{2}exp[-({\varDelta x}^{2}+{\varDelta z}^{2})/{{l}_{c}}^{2}]
\end{align}

Here and following, $\left\langle f(\Delta x,\Delta z) \right\rangle$ represents the ensemble average of the function $f(\Delta x,\Delta z)$ over many different disorder potentials with the same $\sigma$ and $l_c$. Without loss of  generality, in the following we set $l_c=1$ and $\sigma=1$, and tune $d$ from local Kerr limit, $d=0$, to examine the impact of the nonlocality on the branched flows. To enable  a comparison with the branched flow under linear conditions, we additionally conducted simulations where $\sigma=0$. 

The algorithm used to solve Eq.~(1) and (2) is a standard split-step FFT algorithm, and in the simulation, a typical transverse grid size of $\delta x=0.01$, and a stepsize along the light propagation axis of $\delta z=0.01$ were used. Even smaller grid sizes and stepsizes were tested to ensure that the  simulation results were convergent. For the majority of the results presented in the work, the maximum value of nonlocality $d$ was limited to 1, while our computational window spans 50 units. This means  that the nonlocality length $d$ is significantly smaller than the window size. However, in a few instances, we also conducted simulations with larger values of $d$, such as $d=100$. In these cases, we utilized a larger window size of 500 to ensure that all results presented in the work were not affected by the finite-size effect.

\section*{3 Result and discussion}

The simulation results are presented in Fig.~1 and Fig.~2, where a plane light wave with an amplitude $A=0.4$  was assumed to propagate through a disorder potential. We start from the simulation by reproducing the results under linear condition, where $\sigma=0$, and the propagation dynamics is shown in Fig.~1 (a). This plot clearly shows that the plane wave rapidly splits into several channels of enhanced intensity upon propagation, and these channels continue to divide as the wave further propagates. When the self-focusing effect of light wave is introduced into the material, as depicted in Fig.~1(b) for  $\sigma=1, d=0$, the branched flow persists, but now the channels become more concentrated, and thus the self-focusing nonlinear effect promotes the appearance of the wave branching.  To characterize the properties of the branching flow, we introduce two key parameters. The first parameter is the
scintillation index, $S$, defined as:

\begingroup
\setlength{\abovedisplayskip}{6pt} 
\setlength{\belowdisplayskip}{6pt} 
\begin{align}
S(z)=\frac{\left\langle I(x, z)^{2}\right\rangle}{\langle I(x, z)\rangle^{2}}-1
\end{align}
\endgroup

which quantifies the average variance of the intensity distribution. The scintillation index is commonly  used to identify the onset of the first branching point, where it exhibit a peak. The angular brackets in definition of Eq. (4) represent the outcome obtained over many times (100 times in the present study) realizations of the random potential $V(x, z)$ with identical correlation length $l_c$ and strength $\epsilon$. For each realization, $S(z)$ is measured at the specified distance $z$ averaged across the transverse coordinate $x$ (spanning typically  $50 l_c$ in width for each realization). 
The obtained evolution of $S$ as a function of propagation distance $z$ is given in Fig.~2(a). By comparing the location of the peak $S$ between linear and self-focusing regimes, it is evident that the first branching occurs earlier in the self-focusing regime. Furthermore, the overall value of $S$ in the self-focusing condition is notably higher than that in the linear condition, suggesting enhanced intensity of the flows due to the self-focusing effect. This is confirmed by the cross-section distribution of the branched flow, as shown in Fig.~2(c). All these observation agree well with the results reported in previous studies~\cite{mattheakis2016extreme,akhmediev2009extreme,mattheakis2015extreme,zannotti2020caustic,green2019branched,jiang2023nonlinear}.

\begin{figure}[H]
     \centering
     \includegraphics[width=0.8\linewidth]{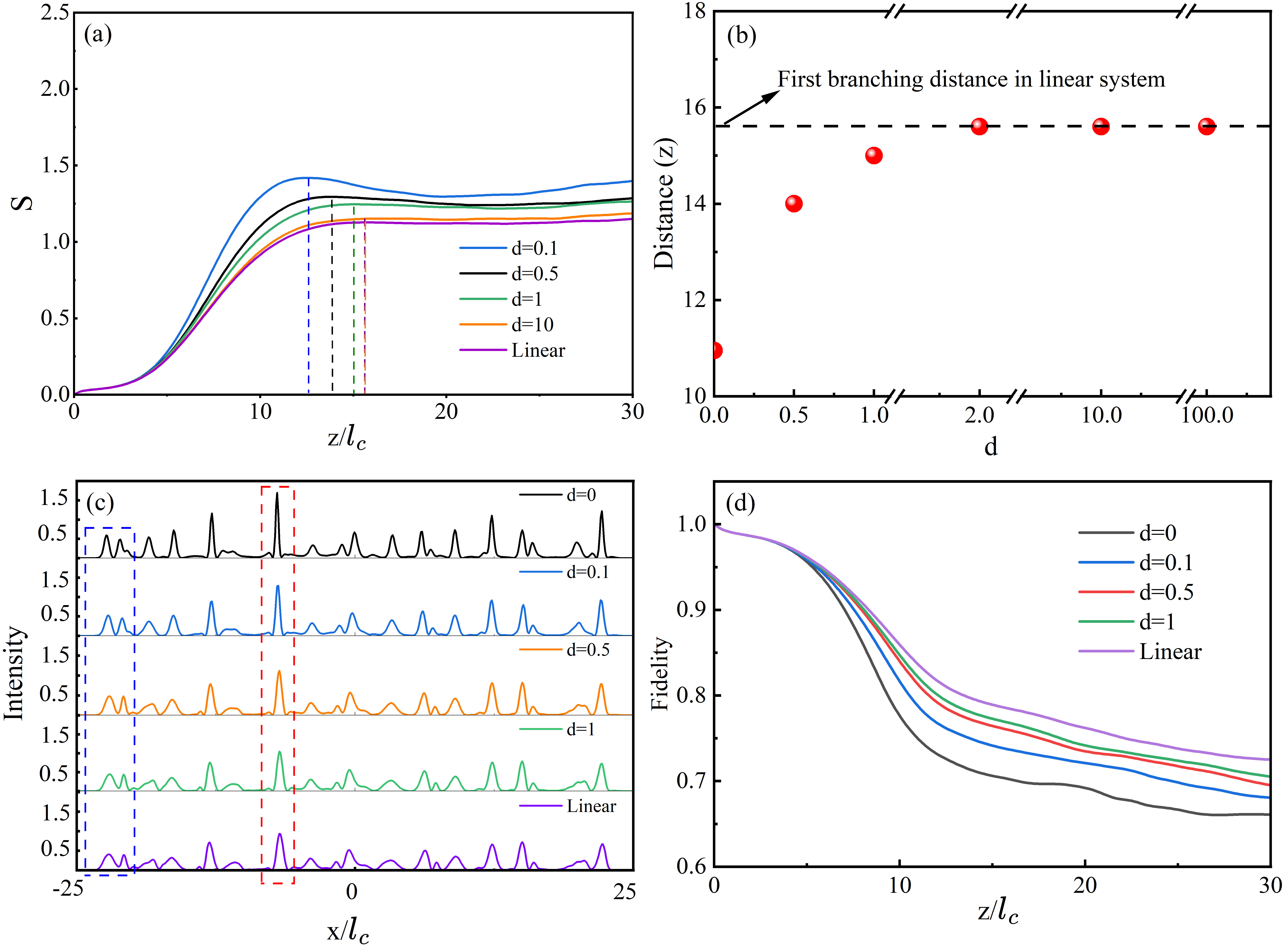}
     \captionsetup{font={stretch=1}}
     \captionsetup{font=footnotesize} 
     \captionsetup{labelfont=bf,name=Fig.}
     \caption{
     {\bf   Comparison of branched flow under varying nonlocal nonlinear conditions.
     } 
     (a) The variance of the scintillation index $S$ as a function of distance $z$. Dashed lines indicate the z-position of the first branching point. (b) The distance at which the first branching occurs. (c) The cross-section distribution of branched flow intensity at respective branching points. (d) The variance of the fidelity $F$ as a function of distance $z$. The results of (a) and (b) presented here are obtained after averaging over 100 specific realizations of disorder potential.
    }
     \label{fig2}
 \end{figure}

The central finding of the present study is that the nonlocality of the nonlinear response tends to conceal the enhanced effects of the local, self-focusing and restores the branched pattern to its linear condition. Indeed, as the degree of nonlocality $d$ increases,  it becomes evident that the first branching point, corresponding to the peak $S$, continuously occurs at greater distance $z$, as demonstrated in a series of propagation dynamics shown in Fig.~1(b-e) and Fig.~2(b) for  $d$  values ranging from $d=0$ to  $d= 0.1, 0.5$,  and until $d=1, 2$. Interestingly enough, when $d=1$,  the light propagation dynamics already aligns well with that observed in the linear condition [ cf. Fig.~1(f) and Fig.~1(b)], and the location of the first branching point nearly coincides with the linear one, though exact coincidence is achieved at $d=2$, see, Fig.~2(b).

Accompanying with the delay in branching, the scintillation index, which rises to a  high level with the pure Kerr effect, drops gradually with  increasing $d$, as illustrated in Fig.~2(a). This implies a reduction in the peak intensities of the branched flows and a blurred out structures. This feature can be attributed to the nonlocality response, which is known acts as a kind of "averaging" effect  over spatial regimes defined by the length $d$.

As a result of the "averaging" effect due to the nonlocality, two typical scenarios for the flow structures are observed with increasing nonlocality $d$, as indicated by two dashed box in Fig.2(c). The first scenario involves the gradual merging of two (or more) channels into one channel as $d$ increases. This scenario is evident in the left dashed box, where two  well-defined  channels at $d=0$ combines into one at $d=0.5$ and beyond. Thus, nonlocality tends to reduce the number of flows. The second scenario involves the smoothing of the intensity of the flow pattern, which is seen in the right dashed box, where  a sharp channel observed at $d=0$ continuously diminishes and broadens, ultimately ending  up with a moderate intensity similar to its neighbours. Both the merging of channels and broadening of channels lead to a weakening of light splittings and, consequently, the suppression of the branching. 

\begin{figure}[H]
     \centering
     \includegraphics[width=1\linewidth]{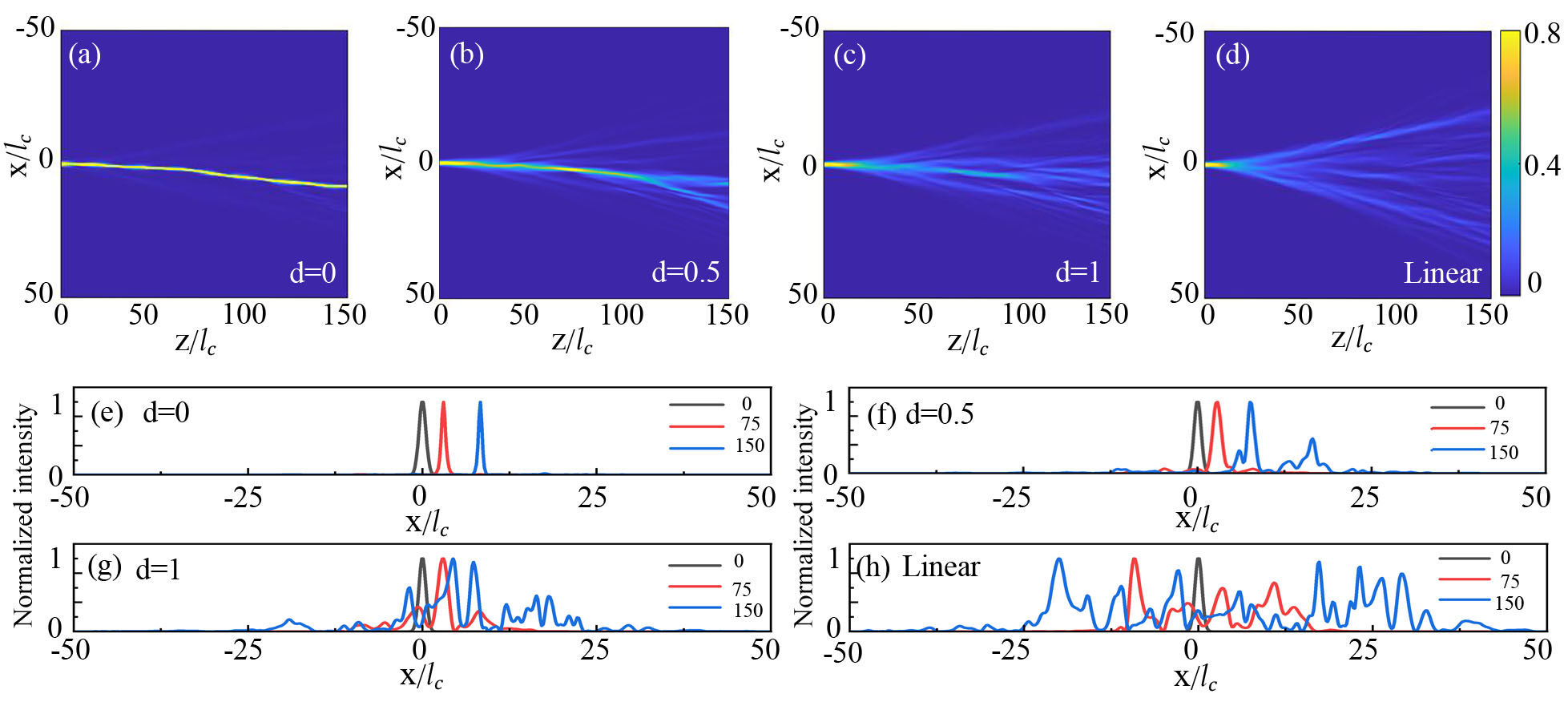}
     \captionsetup{font={stretch=1}}
     \captionsetup{font=footnotesize} 
     \captionsetup{labelfont=bf,name=Fig.}
     \caption{
     {\bf   Propagation of a Gaussian beam through a random potential under various nonlocal nonlinear conditions, with nonlinear coefficient $\sigma$ and degree of nonlocality $d$.
     } 
     (a-c) presents the nonlinear cases with $\sigma=1$  and $d$ values of 0, 0.5, 1, respectively. (d) presents the linear case with $\sigma=1$.
(e-h) The intensity distribution at three  propagation distances $z=0, 75, 100$, represented by black, red, and blue curves, respectively. For improved visibility of the branched flows, the light intensity at each distance has been normalized.
    }
     \label{fig3}
 \end{figure}

Quantitatively, the suppression of the branching can be characterized by the parameter fidelity $F(z)$, which is commonly employed to describe the deformation of light during its propagation. It is defined as folows:

\begin{align}
\label{F}
  F(z)=\frac{\int d x I\left(x,z_{0} \right) I(x, z)}{\sqrt{\left(\int d x I^{2}\left(x, z_{0}\right)\right)\left(\int d x I^{2}(x,z)\right)}}
\end{align}

where $I\left(x, z_{0}\right)$ and $I(x, z)$ represent the intensity distributions in the cross-section at the initial position $ z_{0}$ and at a later position $(z>z_{0})$ along the propagation axis. Here we set $z_0=0$, and the evolution of $F$ with distance $z$ is shown in Fig.~2(d). 

As expected, in all cases, $F$ decreases with $z$ as the waves  continuously experiences scattering by the random potential during propagation. However, a comparison of the fidelity decay curves for different $d$ values reveals that stronger nonlocality corresponds to a slower decline in fidelity, indicating less distortion in wave propagation and, consequently,  weaker branching behaviors.

Finally, we also examine the evolution of a narrow Gaussian beam in the random potential as the influence of nonlocality increases. The Gaussian beam is taken as $\psi(x)|_{z=0}=Ae^{-\frac{x^{2}}{w^{2}}  } $ with an amplitude $A=1.1$ and width $w=1$, and its propagation dynamics is presented in Fig.~3. In the pure self-focusing condition, where $d=0$, the narrow Gaussian beam locally induces a significant change in the refractive index(nonlinear potential). This change, in turn, confines the Gaussian beam, resulting in the formation of a localized light beam, i.e., a spatial optical soliton. As the potential varies randomly with distance, this soliton continues to  deflect, leading to  a self-routing spatial soliton. Please also refer to the light profiles at three distance $z$, which shows the narrow beam maintains its profiles all the time, but appears in different locations(Fig.~3(e)).

As nonlocality $d$ increases, however, it smooths out the underlying nonlinear potential, and consequently, the nonlinear focusing effect becomes not strong enough to confine the Gaussian beams, and it begins to diffract and branch(Fig.~3(b)). Obviously, further increasing $d$ leads to a further weakening of the nonlinear effect. Thus, the branching behavior becomes more profound and, as seen in the plane wave excitation, it eventually approaches the branched flow of a Gaussian beam in the linear condition(cf. Fig.~3(c) and Fig.~3(d)).

Note that such nonlocality-enhanced branching of the narrow beam excitation stands in stark contrast to the soliton dynamics observed in random nonlocal nonlinear media, where, not only is the response of the nonlinear effect nonlocal, but the random potential itself also exhibits nonlocality~\cite{folli2010frustrated, maucher2012stability}. Consequently, as nonlocality averages out the random potential, one observes a suppression of the random walks of solitons induced by nonlocality~\cite{folli2010frustrated}, as well as an enhancement of soliton stability due to nonlocality~\cite{maucher2012stability}.

\section*{4 Theoretical analysis}

Metzger et al. suggested a ray-tracing model to explain the origins of branched flow phenomena\cite{metzger2010branched}, by following the curvature of light described by the differential equation, which, after adapted to (1+1)D system considered in our study, has the following form,

\begin{align}
\label{cl}
  \frac{d}{dz}u(z) +u^{2} (z) =\frac{1}{2} \frac{\partial ^{2} }{\partial x^{2}} V(x,z),
\end{align}
where $u_(x,z ) =\frac{1}{k} \frac{\partial  }{\partial x  }P(x,z )$ is the curvature of light rays, $k$ is wave number, $P(x,z )$ is the phase of the light beams.
Including  the nonlinear change of the refractive index term $n$ to equation (6) yields,

\begin{align}
\label{eq:7}
  \frac{d}{dz}u(z) +u^{2} (z) =\frac{1}{2} \frac{\partial ^{2} }{\partial x^{2}} V(x,z)+\frac{1}{2} \frac{\partial ^{2} }{\partial x^{2}} n(x,z)
\end{align}

Substituting equation(2) into equation (7) one has:
\begin{align}
\label{eq:8}
  \frac{d}{dz}u(z)=-u^{2} +\frac{\sigma }{2} \frac{\partial ^{2} }{\partial x^{2} }I(x,z) +\frac{d}{2} \frac{\partial ^{4} }{\partial x^{4}  }n(x,z) +\frac{1}{2}\frac{\partial ^{2} }{\partial x^{2} } V(x,z)
\end{align}

In equation (6-8) ,  $u(z)\longrightarrow -\infty $, implying that the curvature of the light beam approaches negative infinity, corresponds to the emergence of branching points. The nonlinear and nonlocal terms correspond to the second and third terms on the right-hand side of equation (8), respectively. As evident from the equation, both terms can influence the curvature of the light ray, thus affecting the branched flow; however, their influence on the branch flow are  just opposite, as detailed below.

\begin{figure}[H]
     \centering
     \includegraphics[width=0.6\linewidth]{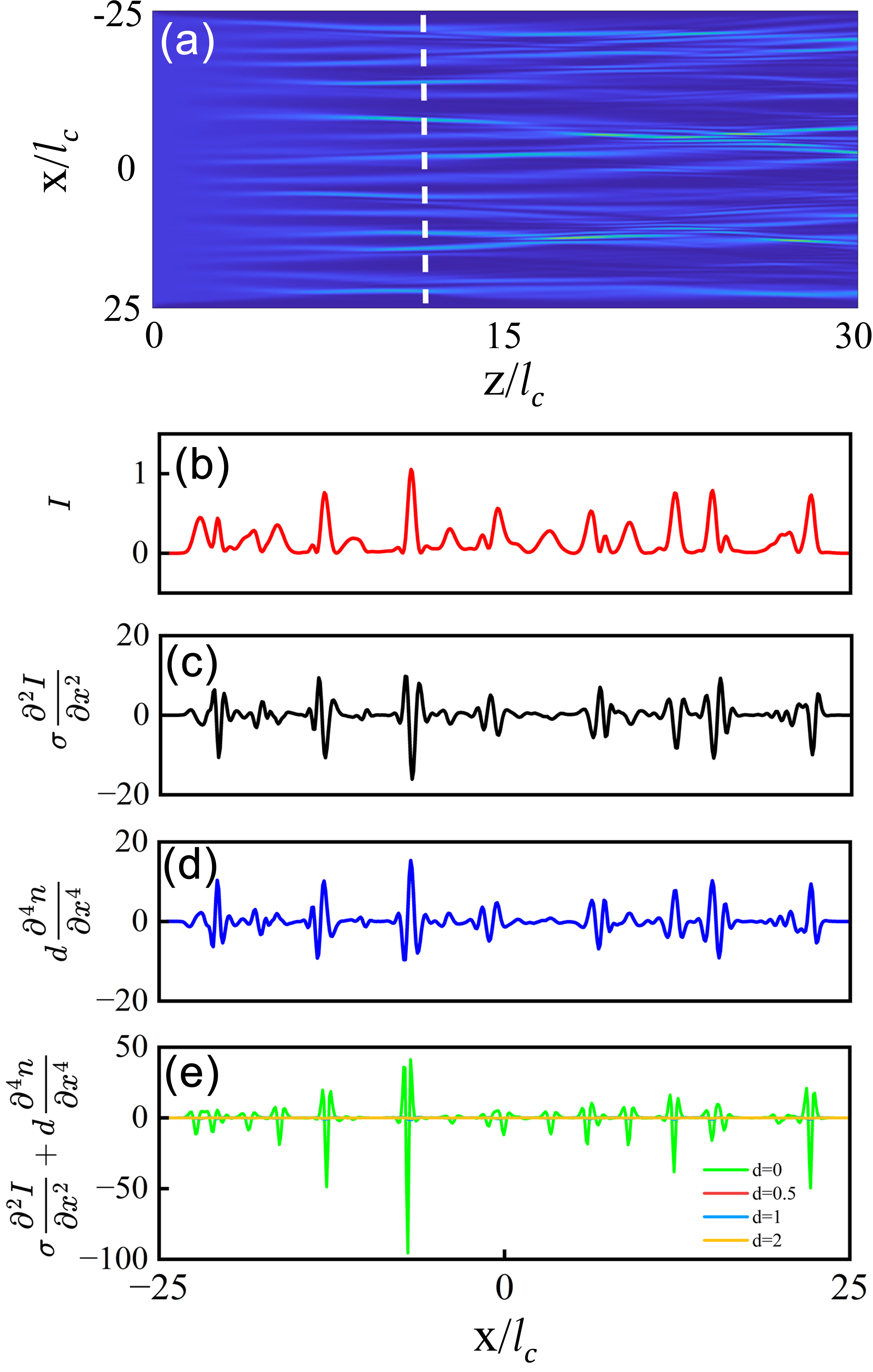}
     \captionsetup{font=footnotesize} 
     \captionsetup{labelfont=bf,name=Fig.}
     \caption{
     {\bf   The ray-tracing model with the nonlocal nonlinear response.
     } 
     (a) The dynamics of branched flow in a nonlocal nonlinear medium, with a degree of nonlocality $d=1$ and a nonlinearity coefficient $\sigma=1$.
 The white dashed line indicate a z-position prior to the occurrence of the first branching point (95\% of the first branching distance).
 (b)–(d) The distributions of light intensity $I$, the Kerr term $\sigma  \frac{\partial ^{2} }{\partial x^{2} }I(x)$,  and the nonlocality term $d \frac{\partial ^{4} }{\partial x^{4}  }n(x)$,  measured at the position corresponding to the white dashed line in (a). Note that there are precisely sign flips in the peaks and dips between the curves in (c) and (d).
 (e) The sum of the nonlocal and Kerr terms for varying $d$ cases, that are all measured at a z-position that is 95\% of the occurrence of the first branching points for each case.
    }
     \label{fig4}
 \end{figure}

In Fig.~4, as a specific example, we present a propagation result, and then show the distributions of $I(x)$ at a selected distance $z$ before the first branching point. We then calculate the curvature contribution from nonlinearity,
$\sigma  \frac{\partial ^{2} }{\partial x^{2} }I(x)$, and from the nonlocality, $d \frac{\partial ^{4} }{\partial x^{4}  }n(x)$, at the selected distance $z$. It is readily seen that the curves of the nonlinearity term  (Fig.~4(c)), at the $x$-positions corresponding to the light intensity peaks (Fig.~4(b)), are always negative, thus, speeding up the curvature u(z) towards negative infinity and leading to the earlier occurrence of the branching point. On the other hand, at those $x$-positions corresponding to the peaks of $I(x)$, the nonlocality curves are always positive, which hinders the curvature u(z) from approaching negative infinity, thus hindering the appearance of the branching point. In other words,  the contribution from the self-focusing nonlinearity is always offset by the contribution of the nonlocality, thus leading to the gradual postponement of the branching.

Eventually, as shown in Fig.~4 (e), when the nonlocality $d$ reaches a sufficiently large value of approximately $2$ and beyond, the two contributions nearly cancel each other out, thereby restoring the branching flow to the linear condition. It is noteworthy that this prediction is in perfect agreement with the numerical results based on the scintillation index $S$, as illustrated in Fig.~2(b), which also demonstrates that the branching point returns to the linear case and saturates for $d \geq 2$.

\section*{5 Conclusion} 

In conclusion, we have investigated the impact of the spatial nonlocality of the nonlinear response of the material on light propagation through a varying random potential. Our findings reveal that the branched flow of light, which is enhanced by the local self-focusing effect, is mitigated by nonlocality. As a result, with increasing nonlocality, the emergence of branched flow is delayed, and the flow pattern broadens. Eventually, nonlocality restores the branched flow to its linear condition.
This study provides a new level of flexibility in controlling and tuning branched flow in nonlinear, randomly varying potentials. By adjusting the degree of nonlocality, one can manipulate the onset and characteristics of branched flow, offering new opportunities for optical applications and technologies.

\section*{Acknowledgments} 

F.Y. acknowledges the support of “Shanghai Jiao Tong University Scientific and Technological Innovation Funds”, and Shanghai Outstanding Academic Leaders Plan (grant no. 20XD1402000). P.W. acknowledges funding from the National Natural Science Foundation (grant no. 12304366) and China Postdoctoral Science Foundation (grant no. BX20230217 and 2023M742295).

\section*{Competing interests} 
The authors declare no competing interests.

\section*{Availability of data and materials}
The data that support the findings of this study are available from the corresponding author upon reasonable request.

\section*{Author contributions} 
Tongxun Zhao and Yudian Wang contributed equally to this work. All authors contributed to the writing, review and editing the manuscript. All authors have accepted responsibility for the entire content of this manuscript and approved its submission.


\end{document}